\documentclass[12pt]{revtex4}
\usepackage[pdftex]{graphicx}
 
\begin{document}

\title{Coherence and Optical Emission from Bilayer Exciton Condensates}
\author{D.W. Snoke}
\address{Department of Physics and Astronomy, University of Pittsburgh, Pittsburgh, PA, 15260, USA}

\begin{abstract}
Experiments aimed at demonstrating Bose-Einstein condensation of excitons in two types  of experiments with bilayer structures (coupled quantum wells) are reviewed, with an emphasis on the basic effects. Bose-Einstein condensation implies the existence of a macroscopic coherence, also known as off-diagonal long-range order, and proposed tests and past claims for coherence in these excitonic systems are discussed. 

\end{abstract}

\maketitle

\section{Introduction. Two Types of Bilayer Excitons}

Condensates of electron pairs have long been known as the basis of superconductivity\cite{leggett}. In superconductors, two spin-1/2 electrons bind into a Cooper pair, which is a boson. These bosons then undergo a type of Bose-Einstein condensation, in which a macroscopic number of the Cooper pairs enter a single wave function (see, e.g., Ref.~\cite{snokebook}, Section 11.2.3). In principle, the same thing should be possible with pair states made of electrons and holes instead of two electrons. In this case the attraction between the fermions is not a phonon-mediated effect, but the Coulomb attraction between particles with positive and negative charge. 

In the past two decades, a number of experiments have focused on demonstrating Bose condensation in this type of system \cite{snokemosk,sciencereview}. One version is the well-known polariton condensates \cite{snokelittlePT}, which have several dramatic effects of the condensation. The polaritons have very short lifetime, however. Although this makes them quite interesting for experiments with optical coupling, it also means that the polariton condensates are never in complete equilibrium. Another set of experiments, which we will review here, has focused on creating excitons that are either stable or metastable. These experiments  are based on semiconductor structures with two coupled, parallel quantum wells. 

\begin{figure}
\begin{center}
\includegraphics[width=0.7\textwidth]{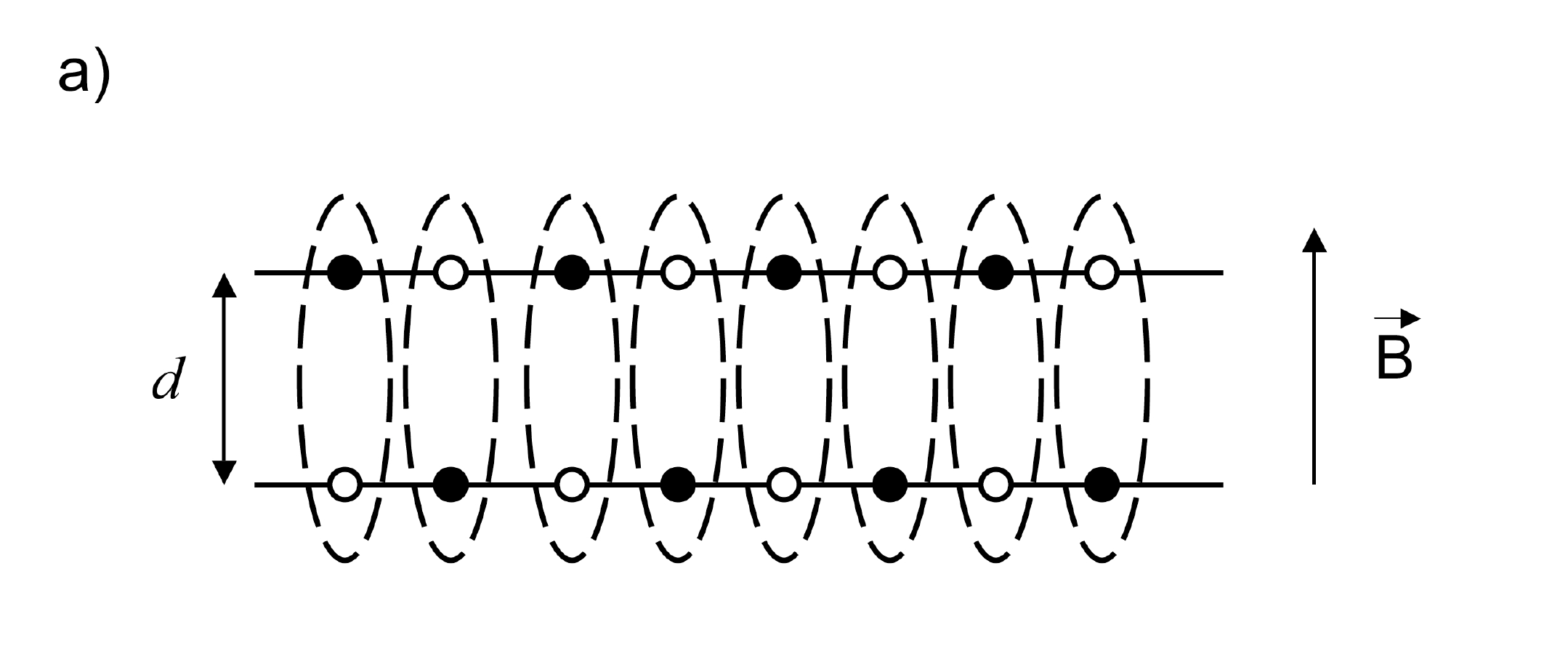}

\includegraphics[width=0.7\textwidth]{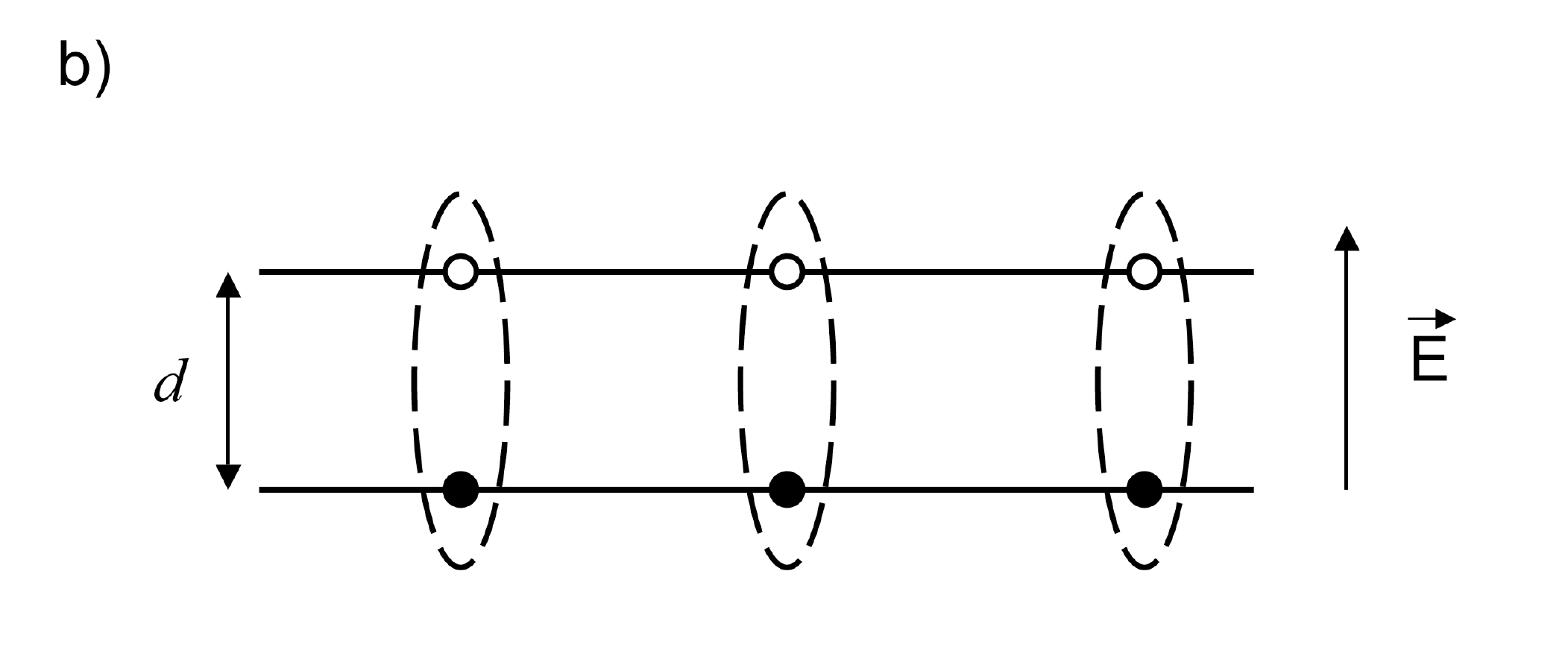}
\caption{a) Schematic of the electrons and holes in a Type A bilayer system. b) Schematic of the electrons and holes in a Type B bilayer system. }
\label{fig1}
\end{center}
\end{figure}

There are two versions of this system, with important similarities and differences. In the first system \cite{eisenstein,shayegan}, which we will call Type A, two quantum wells are created parallel to each other with a thin barrier in between.  Doping is used to create a permanent population of free electrons shared by the two wells. A DC magnetic field is then generated perpendicular to the wells, to create Landau levels. When the magnetic field and free electron density are tuned to the right values, the lowest Landau level in each quantum well will be half filled. One can visualize each of these levels as half full of electrons and half full of holes. Because the wells are near enough to each other for the Coulomb force be significant, the electrons and holes in different wells will be correlated at low temperature.  The state of the system can be described as excitons, as shown in Fig.~\ref{fig1}(a).  In this case, the excitons are not generated optically by taking an electron from the valence band to the conduction band; instead, the excitons in this system are generated by imagining that we start with one well with a full Landau level and the other with a completely empty Landau level, then an electron from the full Landau level tunnels through the barrier to the empty Landau level in the other well. Since the state in which each well has a half-filled Landau level is the ground state of the system, the excitons created this way are stable. 

The second type of system, which we will call Type B, also uses two parallel quantum wells. In this case there is not normally a magnetic field (though one can be added); instead there is an electric field perpendicular to the wells. Free electrons and holes are generated optically and then move into adjacent wells under the influence of the electric field, as shown in Fig.~\ref{fig1}(b). These excitons are known as ``spatially indirect'' excitons, or simply ``indirect'' excitons. In general, all excitons created by optical excitation are metastable and can decay, but if the applied electric field is high enough, the spatially indirect excitons created this way can have very long lifetime, since the lifetime increases exponentially with increasing electric field \cite{szy}. In recent experiments, exciton lifetimes up to 40 $\mu$s have been measured \cite{zoltan-wellwidth,nick}, but much longer lifetime is possible; the main reason for not making the lifetime infinite is to use the optical emission to measure properties of the exciton gas. It is also possible \cite{lilly,lilly2} to generate indirect excitons by injecting holes from a $p$-doped region on one side of the structure and electrons from an $n$-doped region on the opposite side. An analogous system has also been proposed for graphene bilayers \cite{graphene1,graphene2,graphene3}.

There are several important differences between these two types of exciton systems. 

{\bf Exciton Density and Disorder}.  The excitons in the Type B system can be generated with densities ranging from very low to very high, where high density is defined as $na^2 >1$, with $n$ the area density and $a$ the radius of the exciton, i.e., the electron-hole correlation length. The exciton density in this case is directly controlled by the intensity of the pump laser which generates the free electron and holes, or the amount of current, if $p-n$ current generation is used.  

In the low-density limit, the excitons can act as a dilute Bose gas, with the critical temperature for Bose-Einstein condensation increasing with increasing density. In the high-density regime, a condensate of excitons will be in the BCS limit, with critical temperature for condensation decreasing with increasing density. (For a review see Ref.~\cite{ei}; see also Ref.~\cite{snokemosk}, Section 10.3). This limit was studied theoretically already in the 1970's under the name of the ``excitonic insulator.'' It is still proper to talk about bosonic electron-hole pairs and a condensate in this limit, because there is still a weak binding between the electrons and holes.

The excitons in the Type A system are always in this high-density regime. In principle, one could make a dilute exciton gas in this system by having one well keep most of the electrons and the other well have a nearly empty Landau level. The fraction of electrons in each well can be controlled by an electric field perpendicular to the wells. However, the experiments show that when the imbalance is more than a small fraction, the pairing of the carriers is destroyed.  In the BCS-like state with nearly equal numbers of electrons and holes in each well, the binding energy of the excitons in the high-density limit is very weak due to the screening of the Coulomb interaction, which implies very low critical temperature; typical temperatures for these experiments are in the milliKelvin range. 

When disorder in the wells is taken into account, there is an advantage to working in this high density, low temperature regime. Typical disorder fluctuations in the GaAs samples used for these experiments are a fraction of a meV. One meV corresponds to 10 K. If the excitons are at low density and low temperature, they will simply fall into local minima in the disorder potential, below the so-called ``mobility edge.'' In this case they will no longer act as a gas; instead they will act as an ensemble of pinned particles. This limit has been seen in experiments in which the diffusion constant of the excitons in a Type B structure was measured optically; at low density the diffusion constant became nearly zero \cite{zoltan-review}. 

A high density, the electrons and holes will each form a Fermi level that can be larger than the disorder fluctuations. In this case the effect of the disorder potential will be washed out and the electron and hole states will be extended states.  Because the disorder fluctuations are a non-negligible fraction of the exciton Rydberg energy (typical intrinsic exciton binding energies are 4 meV or so in these structures), there is not much room to increase the density above this point before hitting the high-density regime where $na^2 \sim 1$.  As discussed in Section 2.5.4 of Ref.~\cite{snokebook}, the condition $na^2 \sim 1$ is equivalent to the condition $E_F \gg Ry_{ex}$, where $Ry_{ex} = e^2/4\pi\epsilon a$ is the intrinsic two-dimensional exciton binding energy and $a \simeq 4\pi \hbar^2\epsilon/e^2 \mu$ is the intrinsic excitonic Bohr radius, with $\epsilon$ the dielectric permittivity of the medium and $\mu$ the reduced mass of the electron and hole. As noted above, being in the high-density regime does not prevent condensation, but it reduces the critical temperature, as the condensate looks more and more like a BCS state with pair binding energy small compared to the Fermi energy. 

The effect of the disorder is substantially reduced in wider quantum wells \cite{zoltan-wellwidth}, but if the wells are too wide, the intrinsic binding energy of the excitons will be reduced too much, because the binding energy depends on the distance between the electrons and holes \cite{szy}. The optimum well width for GaAs-based structures appears to be around 15 nm. 

{\bf Tunneling current and recombination}. Another significant difference between the Type A and Type B systems is the role of tunneling current. In the Type A system, there is normally no bias normal to the plane of the wells to drive current from one to the other; in-plane electrical bias is used, however, to drive an in-plane current within each well.  In the Type B system, a large electric field normal to the plane of the wells is used to enforce the separation of the electrons and holes into different wells. This will always lead to some current through the barriers.  Tunneling from one well to the other is responsible for the creation of the spatially indirect excitons, but stops once the excitons are formed. The electric field normal to the wells also creates tunneling current into the wells from outside, however, from carriers from the doped substrate and the doped capping layer of the structure. This can be suppressed by various methods, but can still play a large role in the temperature of the excitons, since carriers tunneling in from the substrate and capping layer will be very hot compared to the lattice; up to hundreds of degrees \cite{zoltan-review}. 

A perpendicular magnetic field acts to suppress this tunneling current, since the Landau orbits of the free carriers inhibit them from finding weak spots in the outer barriers to tunnel through \cite{snoke-laikht}. In general, when there is not a magnetic field, the tunneling current through barriers in semiconductor heterostructures is not uniform; since the current is exponentially sensitive to the barrier thickness and alloy content, tiny regions with slightly thinner barrier or lower barrier height will attract most of the current, in ``filaments'' which have been seen dramatically in several experiments \cite{butov-fil}. A magnetic field suppresses  these filaments by forcing the carriers into large Landau orbits which average over the disorder.

\section{Coherence in Exciton Condensates}

\begin{figure}
\begin{center}
\includegraphics[width=0.9\textwidth]{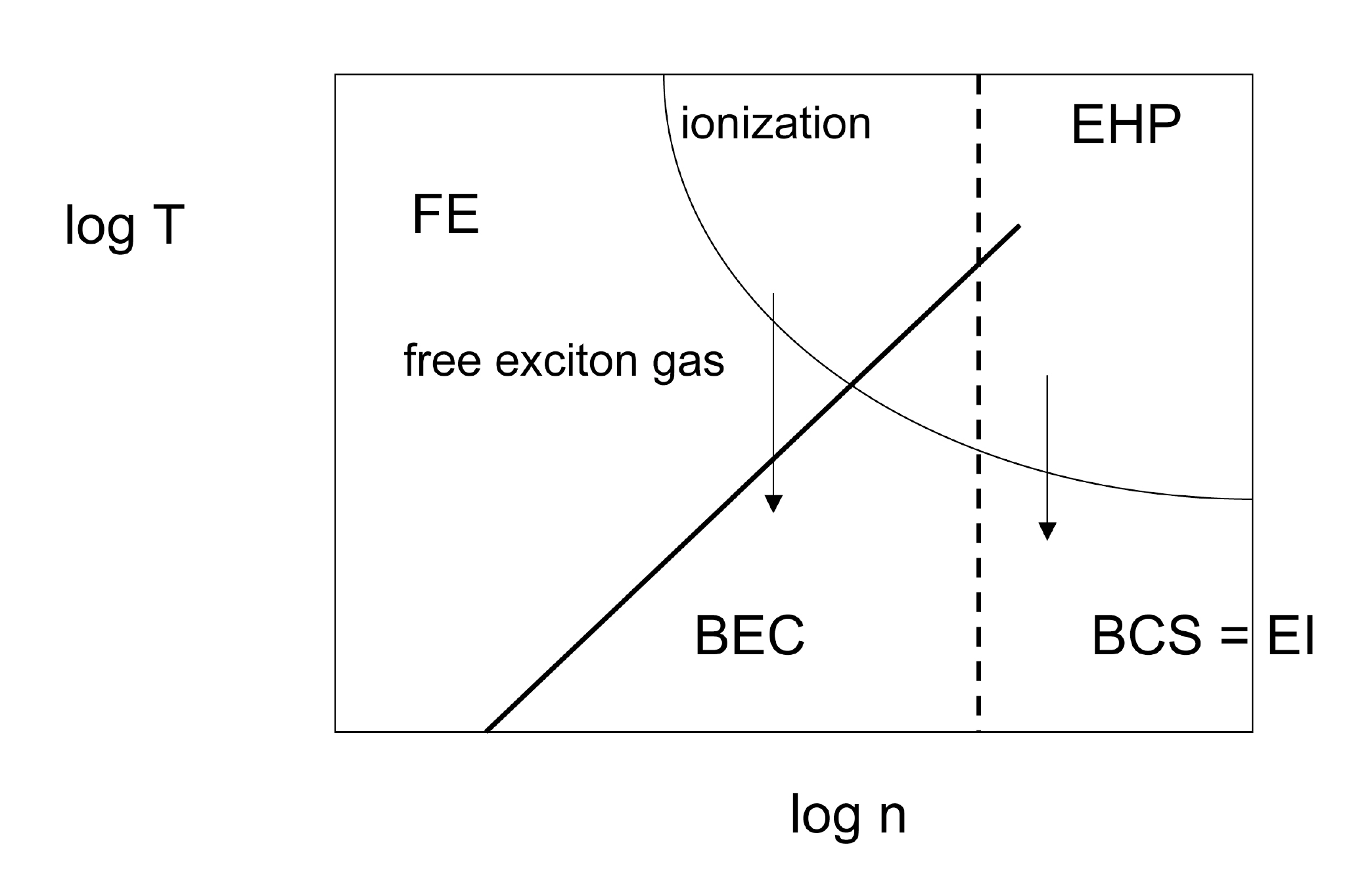}
\caption{General structure of the phase diagram of an electron-hole system. FE = free excitons; EHP = electron-hole plasma; BEC = Bose-Einstein condensate of excitons; EI = exciton insulator state, which is a BCS-like state with a Fermi level for the electrons and for the holes. The transition from BEC to BCS may be gradual. At high pair density $n$, decreasing the temperature gives one critical $T_c$ in which both pairing and condensation occur, in the EI (BCS-like) state. At lower density, there can be two critical temperatures: one for pairing and a second for condensation.}
\label{fig.pt}
\end{center}
\end{figure}

Figure \ref{fig.pt} gives the general structure of the phase diagram for a system with equal number of electrons and holes (or any system with equal numbers of equal-mass, opposite-sign fermions), when disorder is negligible. Since there is actually much confusion about both the terminology and the physics, it is worthwhile to take some time to discuss this phase diagram.
 
The heavy solid line is the phase boundary for BEC in the dilute limit. This is nominally given by the condition that the thermal deBroglie wavelength $\lambda$ of the particles be comparable to the average distance between them, $r_s$, which goes as $n^{-1/2}$ in two dimensions. Since $\lambda \propto \sqrt{\hbar^2/mk_BT}$, the condition $\lambda \sim r_s$ implies that the critical temperature for the transition is proportional to $n$. (Strong interactions and correlations of the excitons can substantially affect this estimate \cite{rap}.) 

A separate transition is the exciton-plasma phase transition, indicated by the thin solid line in Fig.~\ref{fig.pt}.  This is sometimes called the ``Mott'' transition, but it is not the same as the condition $na^2 \sim 1$. The exciton-plasma phase transition, or ionization transition, actually has a quite complicated structure that depends on the details of the exciton-exciton and exciton-free carrier collisions and on the screening of the Coulomb interaction by free carriers \cite{rice,manzke,snoke-craw,snoke-ssc}. (``Ionization'' here refers to dissociation of the electron and hole from each other, not ionization of the underlying atoms.) There is still much debate about the form of this phase boundary. In general terms, we can say that in the high density regime, we expect a critical temperature for ionization, equal to some fraction of the exciton binding energy, while at high temperature, we expect a critical density above which the exciton gas will become disassociated by collisions. At low density and low temperature the gas should be purely free excitons. 

The condition $na^2 = 1$, indicated by the heavy dashed line in Fig.~\ref{fig.pt}, corresponds to the point at which the Fermi statistics of the electrons and holes become important; i.e., the Fermi level becomes larger than the exciton binding energy. At low enough temperature, the $na^2 > 1$ state corresponds to a BCS condensate, as discussed above. 

One crucial fact to gain from this phase diagram is that {\em pairing and condensation are not the same}.  The BCS pair state assumes condensation and coherence (see, e.g., Ref.~\cite{snokebook}, Section 11.2.3), but it is not the only possible paired state to write down.  In a standard superconductor, condensation and pairing occur at the same critical temperature, and for this reason the two different transitions are equated in many people's minds. Looking at the right-hand, high-density side of the phase diagram in Fig.~\ref{fig.pt}, if the carrier density is high, and the temperature is increased from zero, the pairs will unbind at temperatures well below the point at which $\lambda \sim r_s$. This is what happens in a standard BCS superconductor and in the excitonic insulator (EI) phase discussed above: because the binding energy of the pairs is small, pair breaking controls $T_c$. However, in an excitonic system with strong pairing, it is easily possible to have a situation in which the pairs are strongly bound but $\lambda \ll r_s$, as on the left side of the phase diagram of Fig.~\ref{fig.pt}. All exciton gases are comprised of pairs, but these are incoherent pairs, uncorrelated with each other, until they undergo BEC at low temperature. 

At intermediate particle densities, it is possible to have a situation in which there is a plasma-exciton transition to bound pairs (excitons) which are uncorrelated with each other, and then a second transition to a BEC at lower temperature. The transition from electron-hole plasma to excitons may be sharp or may correspond to a gradual increase of the number of exciton pairs as the temperature is decreased, according to the Saha equation \cite{snoke-ssc}. 

In the Coulomb drag experiments \cite{eisenstein,shayegan,lilly2}, there is ample evidence that electron-hole pairs form which move together as excitons do.  The evidence for condensation in those experiments, however, is much more indirect.  A condensate is a coherent state, and therefore evidence of condensation is most directly seen in evidence of coherence, especially interference. This is seen, for example, in the Josephson effect in superconductors, including SQUIDs \cite{snokebook}, in the interference of two condensates in cold atom gases \cite{ketterle}, in measurements of first- and second-order coherence in polariton condensates \cite{deveaud,snoke-science}, and in superradiant Brillouin scattering in magnon condensates \cite{magnon}. It is also evidenced in observation of quantized vortices in liquid helium and in superconductors. 

It is important to understand what we mean by coherence in this context. One type of coherence refers to the phase coherence in the superposition of a single electron and hole in a pair. This is seen, for example, in enhanced tunneling between the two layers \cite{eisenstein-ssc}.  Another type of coherence, the kind which is crucial to the notion of a condensate, is macroscopic coherence of many bosons in the same quantum state. This type of coherence is evidenced by in-plane correlation. For a true condensate, this type of coherence, known as off-diagonal long-range order, will lead to infinite in-plane coherence length. In a translationally invariant two-dimensional system, this cannot occur, and instead a Berezhinskii-Kosterlitz-Thouless transition can occur to a quasicondensate state with coherence correlation falling with a power law dependence \cite{KT}. In a finite, trapped two-dimensional system, e.g. the indirect excitons trapped in harmonic potentials using inhomogenous strain \cite{zoltan-trap}, the coherence length can be comparable to the size of the trap. 

If all the carriers are in excitonic pair states, then there will be enhanced tunneling between the two layers simply due to the pairing, even if the excitons are not correlated with each other in the plane. This effect, which occurs in the Type A bilayer systems \cite{eisenstein-ssc}, is equivalent to the well-known enhancement of the oscillator strength for optical recombination in a Type B bilayer system when electrons and electrons and holes form excitons. 

It is common to interpret many of the results of Coulomb drag and tunneling experiments in Type A systems in terms of a BCS wave function \cite{eisenstein-ssc,bcs1,bcs2}. A BCS wave function is equivalent to a coherent state of bosons (see Ref.~\cite{snokebook}, Section 11.2.3). However, the fact that a BCS wave function can be used to describe some experimental results is not the same as showing coherence; it must also be shown that the results can {\em not} be described by an ensemble of independent pairs. As discussed above, enhanced tunneling and interlayer drag can be explained by electron-hole pairing, without invoking coherence among the pairs. 

It has been argued \cite{eisenstein-private} that experiments with a weak in-plane magnetic field in addition to the large perpendicular magnetic field give evidence of long-range in-plane correlation. In these experiments, when an in-plane magnetic field is applied, the tunneling current is suppressed. A natural in-plane length scale $\lambda$ is defined by setting the flux quantum $\Phi = h/q$ equal to $BA = B(d\lambda)$, where $d$ is the interlayer separation, which implies $\lambda = h/qBd$. 
Experimentally, the measured in-plane $B$ field which suppresses the tunneling implies $\lambda \sim 2 \ \mu$m. This natural length scale can in turn be expressed as an in-plane momentum $\hbar k = h /\lambda = qBd$.  
\begin{figure}
\begin{center}
\includegraphics[width=0.5\textwidth]{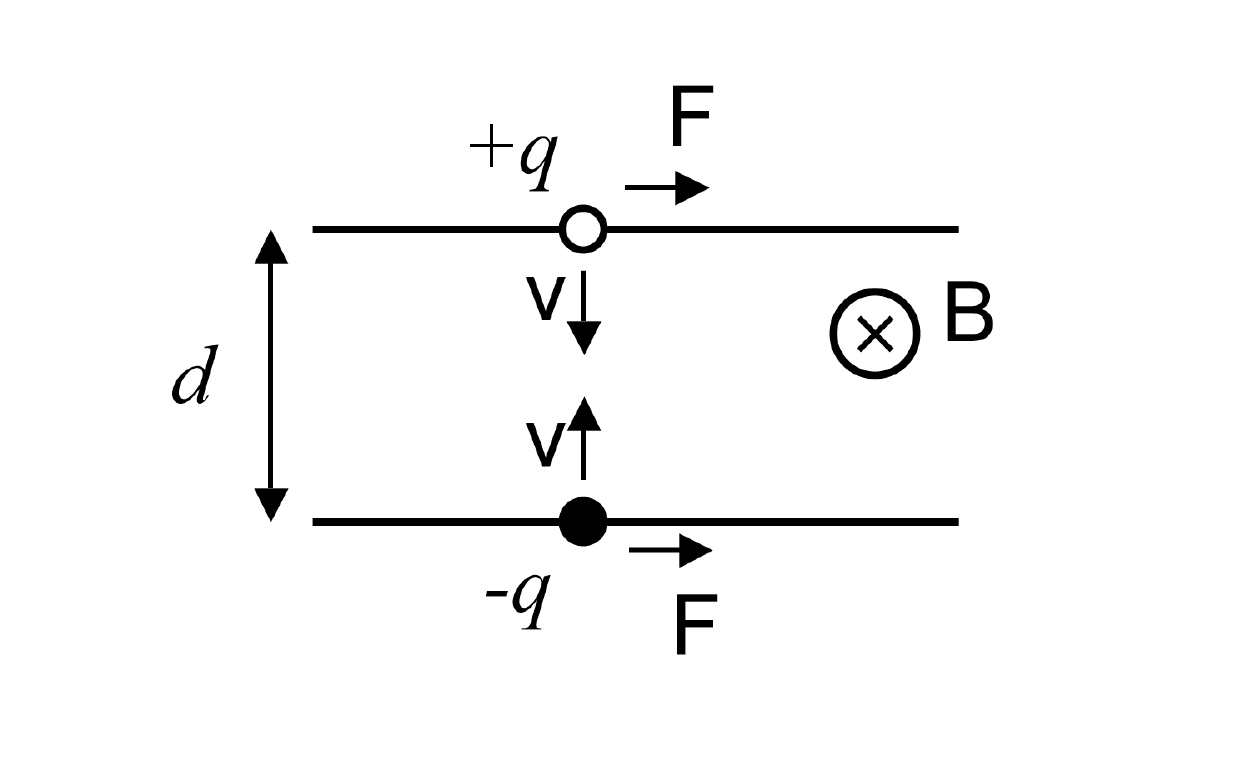}
\caption{Schematic of the effect of a DC in-plane magnetic field on electrons and holes in a bilayer system with tunneling. }
\label{tunnel}
\end{center}
\end{figure}

Figure \ref{tunnel} shows how this momentum comes about. If a carrier tunnels through the barrier with thickness $d$ in a time $\tau$, it will have an effective velocity of $v = d/\tau$. This leads to an in-plane force of magnitude $F = qvB = qdB/\tau$. This force leads to a momentum since $ F = \Delta p/\Delta t$. Since $\Delta t = \tau$ here, we obtain $\Delta p = qdB$. The forces for the negative-charge electrons and positively-charged holes are both in the same direction. 

The in-plane $B$-field therefore introduces a finite momentum to the tunneling carriers, and the suppression of the tunneling current can be viewed as a measurement of a critical velocity of the excitons. One explanation for this may be related to the Landau critical velocity for superfluids (e.g., Ref.~\cite{snokebook}, Section 11.1.4). Other effects may lead to a critical velocity, such as ionization of the weakly-bound excitons due to collisions with defects in the lattice. The fact that no critical temperature has been measured for the bilayer resonant tunneling effect in Type A structures makes it difficult to assess whether it is related to condensation.



\section{Optical Signatures of Exciton Coherence}

In the case of Type B excitons, the observation of coherence is expected to be much easier, but experimental results so far have not provided definitive evidence. 

Since the excitons in this type of system couple directly to photons that leave the system, the coherence of the excitons should map directly to optical coherence of the emitted photons, and the in-plane coherence length can be measured by first-order interference. This has been done with polariton condensates \cite{deveaud,snoke-science}, but with indirect excitons in Type B structures, the evidence is not as clear. Butov and coworkers \cite{butov} have reported interference of light emitted from indirect excitons in a Type B structure at low temperature. The spectral width of this emission was around 2 meV, which by the W-K theorem (see, e.g., Ref.~\cite{snokebook}, Section 9.6) implies a coherence time of about 300 femtoseconds, much less than the lifetime of the excitons, and not too different from what would be expected for incoherent light from a small source, although there was clearly an enhancement of the coherence in these experiments as the temperature decreased. 

Coherent light emission alone is not a sufficient test for exciton condensation, because lasers also emit coherent light, and laser light comes from unpaired, incoherent electrons and holes. To show that the electrons and holes make up a coherent excitonic condensate, it is also necessary to show a) that there is long-range in-plane correlation of the coherent emission, and b) that the excitons are still good quantum states and are not strongly dephased. Both of these have been shown in the polariton systems, in which two transitions, one for condensation and a second, at higher pair density, for lasing have been demonstrated \cite{2thres}. 

\begin{figure}
\begin{center}
\includegraphics[width=0.5\textwidth]{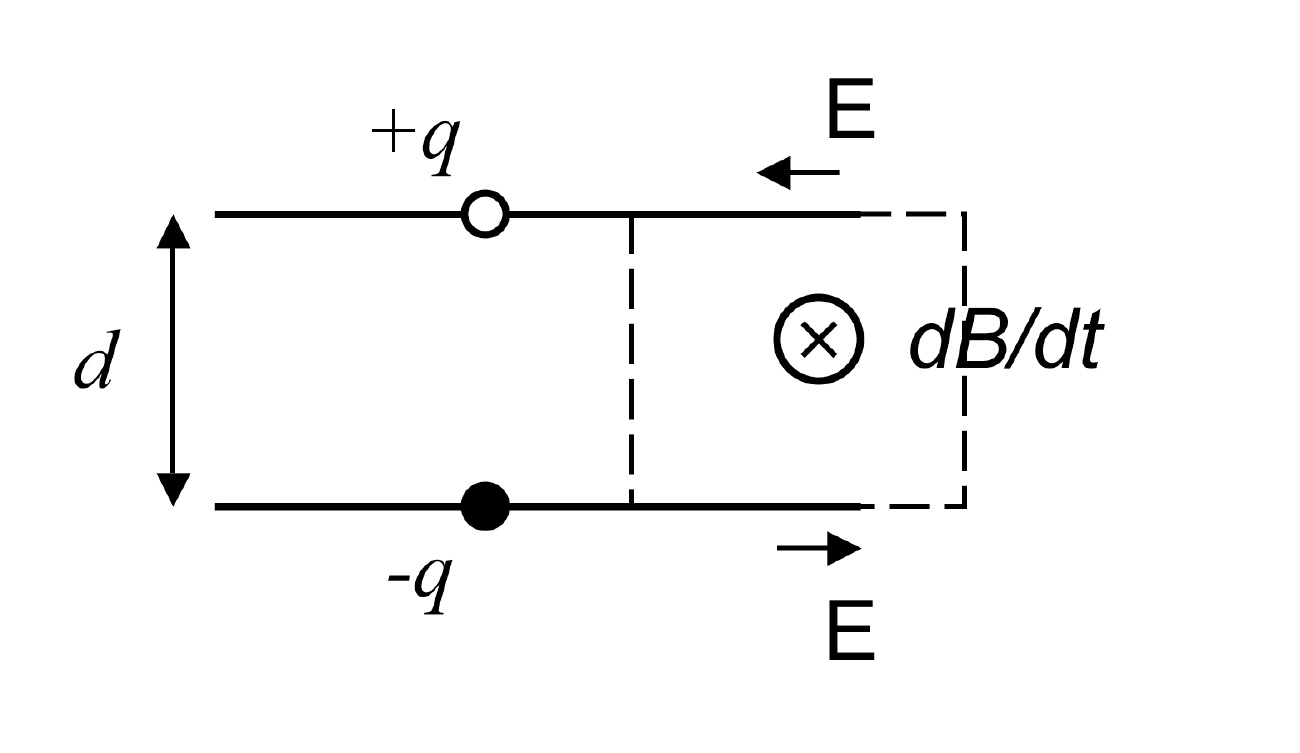}
\caption{Schematic of the effect of an AC in-plane magnetic field on electrons and holes in a bilayer system with no tunneling. }
\label{littlemag}
\end{center}
\end{figure}

Balatsky, Joglekar, and Littlewood  \cite{little-inplane} proposed an interesting test for the presence of a condensate of indirect excitons in a Type B structure.  Similar to the experiment represented in Fig.~\ref{littlemag}, an in-plane magnetic field is used, but this time an AC, time-varying field. By Maxwell's equation $\nabla\times \vec{E} = -\partial \vec{B}/\partial t$, this implies a time-varying difference of the in-plane component of the electric field between the two spatially separated planes. (Perpendicular electric field can be eliminated by symmetry if the in-plane magnetic field is spatially homogeneous.) Since the electron and hole have opposite charge, this leads to a net in-plane force on the excitons. 

If the excitons are incoherent, then their motion under this force should follow an Ohm's law-type drift behavior in which their velocity is proportional to the force $qE$.  On the other hand, if the excitons are superfluid, then they should have an acceleration proportional to the force. If the magnetic field oscillates sinusoidally, then the superfluid component should oscillate in phase with the magnetic field while the normal component oscillates out of phase. This could, in principle, be measured by optical imaging of the exciton motion. 

The magnitude of the oscillation is estimated by the following calculation. The electric field magnitude is $E_0 = \omega B_0 d$, and the maximum acceleration of the exciton is is related to the maximum amplitude of the spatial motion, $x_0$, by $a_0 = \omega^2 x_0 = qE_0/m$. This implies
\begin{equation}
x_0 = \frac{q}{m}\frac{B_0d}{\omega}. 
\end{equation}
If we want $x_0$ to be around 10 $\mu$m, to be resolvable by standard far-field optical imaging, then for  AC oscillation frequency of 100 kHz (to match the typical exciton lifetime), $d \sim 10$ nm, and mass of the order of typical effective masses of carriers, about 1/10 of a vacuum electron mass, this implies $B_0 \sim 1$ T.  It is probably experimentally too difficult to have a magnetic field of this magnitude oscillate at 100 kHz, but if the spatial resolution of the imaging can be reduced, the required magnitude of the magnetic field can also be reduced. Alternatively, the frequency of the oscillation can be reduced by making the exciton lifetime longer, which will decrease the optical emission signal used to detect the motion. 

This proposal was greeted with some controversy in the transport community because of questions about the current \cite{macdonaldye}. If there is zero tunneling between the two layers, then a circuit cannot be completed and no net current can flow. In the context of the indirect excitons, this means that if the exciton density is constant, then excitons will pile up at one end of the structure and repel other excitons from moving. However, in typical experiments with Type B structures, the exciton density is not constant, and the region over which the excitons are created and the amplitude of the spatial oscillation due to the time-varying field are both small compared to the size of the whole structure. 
There will therefore be no back EMF produced by the lack of completion of the circuit, and the excitons will move essentially the same as if they were single excitons in an empty, infinite two-dimensional space. 

Of course, the motion of the excitons is measured by looking at the photons emitted by them when they recombine, which involves tunneling through the barrier. But this lifetime can be made arbitrarily long, as discussed above, so that this current is negligible. In optical experiments, the photon emission functions as a probe in which only a tiny fraction of the excitons are destroyed at any time.  We can therefore ignore the current due to exciton recombination leading to photon emission in these experiments. 

\section{Conclusions}

There is still much work to be done to demonstrate condensation in bilayer systems. There is no question that there is pairing in both Type A and Type B systems, which leads to enhanced tunneling through the barrier between the wells. But off-diagonal long-range order in the plane of the wells has not yet been clearly demonstrated in either system with a direct test. 

Probably the best bet for doing a direct test of coherence in the Type A bilayer systems would be a transport experiment analogous to a Josephson junction. Two regions with bilayer condensate could be separated by a tunneling junction, and the tunneling current through this in-plane barrier could be measured. 

For the indirect excitons in Type B structures, the proposal of Littlewood and coworkers to use a time-varying, in-plane magnetic field to see undamped superfluid exciton motion is reasonable, but requires high magnetic fields or very tiny spatial resolution which may be unobtainable. First-order coherence in the optical emission is also a test of coherence of the exciton condensate, but requires additional tests to distinguish it from standard lasing and to distinguish it from interference from an incoherent source analogous to stellar interferometry.

{\bf Acknowledgements}. This work has been supported by the Department of Energy through project DOE-DE-FG02-99ER45780. We thank J.P. Eisenstein and P.B. Littlewood for helpful discussions.

\end{document}